\documentclass[reprint,
superscriptaddress,amsmath,amssymb,aps, longbibliography
]{revtex4-2}

\usepackage{graphicx}
\usepackage{dcolumn}
\usepackage{bm}
\usepackage{siunitx}
\usepackage[noend]{algorithmic}
\usepackage{algorithm}
\usepackage[skins]{tcolorbox}
\usepackage{setspace}
\usepackage{blkarray}

\usepackage{xcolor}
\usepackage{soul}
\usepackage{systeme}
\usepackage{footmisc}

\usepackage{scalerel} 

\makeatletter
\def\maketag@@@#1{\hbox{\m@th\normalfont\normalsize#1}}
\makeatother

\usepackage{booktabs}
\usepackage{hyperref}
\usepackage[capitalize]{cleveref}
\usepackage[utf8]{inputenc}

\usepackage[lining,semibold]{libertine} 
\usepackage{amsthm}
\usepackage[libertine, cmintegrals, bigdelims, vvarbb]{newtxmath}

\usepackage{amsmath}
\usepackage{amsfonts}
\usepackage{mathrsfs}
\usepackage{gensymb}
\usepackage{bbm}
\usepackage{dsfont}

\usepackage{pgfplots}
\usepgfplotslibrary{colormaps}

\usepackage{kbordermatrix}
\renewcommand{\kbldelim}{(}
\renewcommand{\kbrdelim}{)}

\usepackage{chemformula}
\usepackage[caption=false]{subfig}

\usepackage{soul}
\usepackage{xcolor}

\theoremstyle{definition}

\usepackage{scalerel} 

\definecolor{webgreen}{rgb}{0,.5,0}
\definecolor{webbrown}{rgb}{.6,0,0}
\definecolor{grigio}{rgb}{.85,.85,.85} 
\definecolor{RoyalBlue}{rgb}{0.0, 0.14, 0.4}
\definecolor{skyblue1}{rgb}{0.45,0.62,0.81}
\definecolor{skyblue2}{rgb}{0.2,0.39,0.64}
\definecolor{skyblue3}{rgb}{0.13,0.29,0.53}
\definecolor{scarlet1}{rgb}{0.93,0.16,0.16}
\definecolor{scarlet2}{rgb}{0.8,0,0}
\definecolor{scarlet3}{rgb}{0.64,0,0}

\definecolor{g}{gray}{0.50}

\hypersetup{%
    colorlinks=true, linktocpage=true, pdfstartpage=1, pdfstartview=FitV,%
    breaklinks=true, pdfpagemode=UseNone, pageanchor=true, pdfpagemode=UseOutlines,%
    plainpages=false, bookmarksnumbered, bookmarksopen=true, bookmarksopenlevel=1,%
    hypertexnames=true, pdfhighlight=/O,
    urlcolor=webbrown, linkcolor=RoyalBlue, citecolor=webgreen, 
    pdftitle={},%
    pdfauthor={Timur Aslyamov},%
    pdfsubject={},%
    pdfkeywords={},%
    pdfcreator={pdfLaTeX},%
    pdfproducer={LaTeX REVTeX}%
}

\begin{document}
\title{Dissipation bounds precision of current response to kinetic perturbations}

\author{Krzysztof Ptaszy\'{n}ski}
\email{krzysztof.ptaszynski@ifmpan.poznan.pl}
\affiliation{Department of Physics and Materials
Science, University of Luxembourg, L-1511 Luxembourg City, Luxembourg}
\affiliation{Institute of Molecular Physics, Polish Academy of Sciences, Mariana Smoluchowskiego 17, 60-179 Pozna\'{n}, Poland}

\author{Timur Aslyamov}
\email{timur.aslyamov@uni.lu}
\affiliation{Department of Physics and Materials
Science, University of Luxembourg, L-1511 Luxembourg City, Luxembourg}

\author{Massimiliano Esposito}
\email{massimiliano.esposito@uni.lu}
\affiliation{Department of Physics and Materials
Science, University of Luxembourg, L-1511 Luxembourg City, Luxembourg}

\date{\today}

\begin{abstract}
The precision of currents in Markov networks is bounded by dissipation via the so-called thermodynamic uncertainty relation (TUR). In our work, we demonstrate a similar inequality that bounds the precision of the static current response to perturbations of kinetic barriers. Perturbations of such type, which affect only the system kinetics but not the thermodynamic forces, are highly important in biochemistry and nanoelectronics. We prove that our inequality cannot be derived from the standard TUR. Instead, it implies the standard TUR and provides an even tighter bound for dissipation. We also provide a procedure for obtaining the optimal response precision for a given model.

\end{abstract}
\maketitle

\textit{Introduction.---}Among the most fundamental results of statistical physics are the relations between the system response to external perturbations and stationary thermodynamic observables. Close to equilibrium, such a relation is given by the seminal fluctuation-dissipation theorem (FDT) linking the linear response to external forces and equilibrium fluctuations~\cite{kubo1966fluctuation}. The link between dissipation and fluctuations has recently been generalized to a far-from-equilibrium regime \cite{agarwal1972fluctuation,seifert2010fluctuation,prost2009generalized,altaner2016fluctuation,chun2021nonequilibrium,baiesi2009fluctuations,dechant2020fluctuation,di2018kinetic,falasco2022beyond,hasegawa2024thermodynamic}. 
In particular, for Markov jump processes, the theory of stochastic thermodynamics~\cite{seifert2012stochastic} gave rise to the thermodynamics uncertainty relation (TUR) linking the entropy production rate $\dot{\sigma}$ which measures dissipation to the average of any current $\mathcal{J}$ and its variance $\langle\langle\mathcal{J}\rangle\rangle$ as~\cite{barato2015thermodynamic, gingrich2016dissipation, pietzonka2016universal, pietzonka2017finite, horowitz2017proof, falasco2020unifying, horowitz2020thermodynamic, vu2020entropy}
\begin{align}
    \label{eq:tur}
    \frac{\mathcal{J}^2}{\langle\langle\mathcal{J}\rangle\rangle}\leq \frac{\dot{\sigma}}{2}\,.
\end{align}
This relation shows that to reduce the fluctuations, one needs to pay by increased dissipation. In other words, it describes the thermodynamic cost of current precision $\mathcal{J}^2/\langle \langle \mathcal{J} \rangle \rangle$. It can also be used to infer the value of entropy production, which is often directly inaccessible, using measurable currents and their fluctuations~\cite{seifert2019stochastic}. Indeed, the current fluctuations have  been measured in various setups, such as nanoelectronic devices~\cite{birk1995shot, gustavsson2006counting} and biomolecular systems~\cite{svoboda1994fluctuation, visscher1999single, bormuth2009protein}, and experimental data have been used to infer the value of the entropy production in the kinesin molecular motor~\cite{seifert2018stochastic}.

In our work, we get closer to the original formulation of FDT by providing a thermodynamic trade-off between the current fluctuations and the static linear current response, rather than the current itself. Recently, static responses for Markov jump processes and chemical reaction networks have attracted notable attention~\cite{owen2020universal, owen2023size, gabriela2023topologically, chun2023trade, aslyamov2024nonequilibrium, aslyamov2024general, harunari2024mutual, floyd2024learning, zheng2024information} due to their importance in biophysical applications (proofreading, sensing \cite{murugan2014discriminatory,owen2020universal}). Specifically, we focus on the response to kinetic perturbations, i.e., perturbations affecting only the kinetics of transitions between the system states but not the thermodynamic forces. Perturbations of this type are crucial in chemistry (including biochemistry), as they correspond to the control of kinetic rates by changing the concentration of a catalyst (e.g., enzyme) \cite{Wachtel_2018}. Among others, kinetics may play a crucial role in determining the direction of motion of molecular motors~\cite{maes2015kinetics} and providing the driving mechanism of Maxwell demons~\cite{esposito2012stochastic}. Chemical kinetics can also be controlled by magnetic fields, e.g., via the radical pair mechanism~\cite{steiner1989magnetic}, which is hypothesized to be the basis for magnetoreception and other magnetic field effects in biology~\cite{ritz2000model, hore2016radical, wiltschko2019magnetoreception, zadeh2022magnetic}. Another important example of kinetic perturbations appears in the field of nanoelectronics, where it corresponds to the adjustment of tunnel barriers in quantum dots~\cite{gustavsson2006counting,gustavsson2009electron,sanchez2019Autonomous} or potential barriers in CMOS devices~\cite{freitas2021stochastic, gopal2022Large} by gate voltages. Crucially, the response of currents to kinetic perturbations vanishes at equilibrium, where the state of the system depends only on its thermodynamics, and not on its kinetics. This leads to the intuitive expectation that the response to kinetic perturbations requires some thermodynamic cost. Indeed, certain thermodynamics bounds on the response to kinetic perturbations have been obtained for Markov jump processes~\cite{owen2020universal, gabriela2023topologically}, chemical reaction networks~\cite{chun2023trade}, and nonequilibrium diffusion~\cite{gao2022thermodynamic, gao2024thermodynamic}. However, they apply to static system observables and their fluctuations rather than to currents.

In this Letter, we provide numerical evidence to conjecture a new bound on the current response to kinetic perturbations. We call it the response thermodynamic uncertainty relation (R-TUR). We can prove it in special cases detailed below, e.g., for unicyclic networks, linear response regime close to equilibrium, and for single edge perturbations. To formulate that bound, we parameterize the transition rates of the opposite Markov jumps $\pm e$ as
\begin{align}
\label{eq:rates-model}
    W_{\pm e}=\exp\big[B_e(\varepsilon)\pm S_e/2\big]\,,
\end{align}
where $B_e$ and $S_e$ parameterize the symmetric and asymmetric part of the transition rate, respectively. The former term characterizes the kinetic barriers, while the latter is the entropy change in the reservoir due to the jump $e$. Only the kinetic part $B_e$ is assumed to depend on the control parameter $\varepsilon$ (which can correspond, e.g., to the enzyme concentration~\cite{Wachtel_2018} or the gate voltage~\cite{gustavsson2006counting,gustavsson2009electron,sanchez2019Autonomous,freitas2021stochastic, gopal2022Large}). The bound reads
\begin{align}
    \label{eq:response-tur}
    \frac{( d_\varepsilon\mathcal{J})^2}{\langle\langle\mathcal{J}\rangle\rangle}\leq \frac{b_\text{max}^2\dot{\sigma}}{2}\,,
\end{align}
where $d_\varepsilon\mathcal{J}\equiv d\mathcal{J}/d \varepsilon$ is the static response of any current $\mathcal{J}$ to the parameter $\varepsilon$ and $b_\text{max}=\max_e |\partial_\varepsilon B_e|$ is the maximal rate of change of kinetic barriers in the system. If we perturb a single barrier corresponding to a given edge $e$, our bound \eqref{eq:response-tur} simplifies to
\begin{align}
    \label{eq:response-tur_Simplif}
    \frac{(d_{B_e} \mathcal{J})^2}{\langle\langle\mathcal{J}\rangle\rangle} \leq \frac{\dot{\sigma}}{2}\,.
\end{align}

Although the above bounds look similar to TUR~\eqref{eq:tur}, we prove that they cannot be derived from TUR. 
Instead, TUR is a consequence of our result. Moreover, we develop the optimization procedure that transforms R-TUR into a bound for entropy production that is tighter than the optimized TUR~\cite{vu2020entropy}. However, in some specific cases R-TUR can be derived from TUR. In particular, \cref{eq:response-tur} can be proven for unicyclic networks and the linear response regime close to equilibrium. Furthermore, \cref{eq:response-tur_Simplif} can be derived when the current response at the perturbed edge is considered, i.e., when $\mathcal{J}=j_e$. 

\textit{Framework.---}We consider a continuous-time Markov process describing stochastic jumps between $N$ discrete states of the system (corresponding, e.g., to chemical configurations, charge states of nanoelectronic devices, or molecule conformations). It is described by the directed graph $\mathcal{G}$, where the nodes of $\mathcal{G}$ correspond to the system states and the edges $\pm e\in\mathcal{E}$ to transitions between states. We focus on the steady state $\boldsymbol{\pi}$ given by the condition
\begin{align}
\label{eq:NESS}
    \mathbb{W}\cdot\boldsymbol{\pi} = 0\,,
\end{align}
where $\boldsymbol{\pi}=(\dots,\pi_i,\dots)^\intercal$ is the vector of state probabilities $\pi_i$ (with $\sum_i\pi_i=1$). $\mathbb{W}$ is here the rate matrix with non-diagonal elements $W_{t(\pm e)s(\pm e)}=W_{\pm e}(\varepsilon)$ describing the transition rate from the state $s(\pm e)$ (the source of the edge $\pm e$) to the state $t(\pm e)$ (the tip of the edge $\pm e$) and with the diagonal elements $W_{ii}=-\sum_{j \neq i}W_{ji}$.
\begin{figure}
    \centering
    \includegraphics[width=0.5\textwidth]{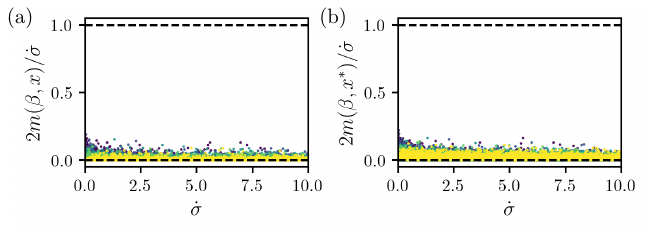}
    \caption{(a): The colored dots are $m(\boldsymbol{\beta},\boldsymbol{x})$ numerically calculated for random values of the vectors $\boldsymbol{\beta}$ and $\boldsymbol{x}$. (b): The colored dots are the optimized $m(\boldsymbol{\beta},\boldsymbol{x}^*)$ with $\boldsymbol{x}^*$ from \cref{eq:opt-x}. The dashed lines in (a,b) denote the bound~\eqref{eq:response-tur}.
    Simulations: we generate graphs with random topology with $N=5$ nodes and various numbers of edges (from $6$ to $10$ highlighted by colors purple, blue, green, light green and yellow). We calculate \cref{eq:rates-model} with randomly distributed $-2\leq B_e\leq 2$ and $0\leq \exp(S_e/2) \leq 10$. For the vectors $\boldsymbol{\beta}$ and $\boldsymbol{x}$ we use $-1 \leq \beta_e\leq 1$ and $-2\leq x_e \leq 2$. For the simulation methods, see \cref{sec:numerical}
    }
    \label{fig:fig-1}
\end{figure}

The bound \eqref{eq:response-tur} can be applied to an arbitrary current observable, namely, $\mathcal{J}=\sum_{e}x_e j_e$, where $j_e = W_{+e}\pi_{s(+e)}-W_{-e}\pi_{s(-e)}$ are the edge currents and $\boldsymbol{x}=(\dots, x_e, \dots)^\intercal$ is an arbitrary vector. Using the chain rule for the static response of $\mathcal{J}$ we arrive at 
\begin{align}
\label{eq:response-chain-rule}
    d_\varepsilon \mathcal{J} &= \sum_e x_e d_\varepsilon j_e = \sum_e x_e \sum_{e'} (d_{B_{e'}}j_{e})\partial_{\varepsilon} B_{e'}\nonumber\\
    &=\sum_{e'}b_{e'} d_{B_{e'}}\mathcal{J}=\boldsymbol{b}^\intercal\nabla\mathcal{J}\,,
\end{align}
where we introduce $\boldsymbol{b} =(\dots, \partial_\varepsilon B_e,\dots)^\intercal$ known from the model parameterization and $\nabla\mathcal{J}=(\dots, d_{B_e}\mathcal{J}, \dots)^\intercal$ where $d_{B_{e'}}\mathcal{J} = \sum_e x_e d_{B_{e'}} j_e$ is the static response to the edge parameters $B_{e'}$. The entropy production rate can be calculated as $\dot{\sigma}=\sum_e j_e S_e$~\cite{seifert2012stochastic}.

\textit{Numerical evidence.---}We start by providing numerical evidence for our bound. Let us first rewrite \cref{eq:response-tur} by expressing $d_\varepsilon \mathcal{J}$ via \cref{eq:response-chain-rule} and using the expression $\langle\langle\mathcal{J}\rangle\rangle=\boldsymbol{x}^\intercal\mathbb{C}\boldsymbol{x}$~\cite{vu2020entropy}, with the covariance matrix $\mathbb{C}$ of the edge currents defined in \cref{sec:numerical}. We obtain
\begin{align}
    \label{eq:response-tur-Markov}
  \forall_{\boldsymbol{\beta}, \boldsymbol{x}}:   m(\boldsymbol{\beta},\boldsymbol{x})
    \equiv \frac{( d_\varepsilon\mathcal{J})^2}{\langle\langle\mathcal{J}\rangle\rangle b_\text{max}^2}
    = \frac{\big(\boldsymbol{\beta}\nabla\mathcal{J}\big)^2}{\boldsymbol{x}^\intercal\mathbb{C}\boldsymbol{x}}&\leq \frac{\dot{\sigma}}{2}\,,
\end{align}
where $\boldsymbol{\beta}=\boldsymbol{b}/b_\text{max}$ with $b_\text{max}=\max_e |b_e|$ and $|\beta_e|\leq 1$.  
From \cref{fig:fig-1}a one can see that the bound \eqref{eq:response-tur-Markov} with random vectors $\boldsymbol{\beta}$ and $\boldsymbol{x}$ indeed holds, but is very loose. Thus, we now look for $\boldsymbol{\beta}$ and $\boldsymbol{x}$ that maximize $m(\boldsymbol{\beta},\boldsymbol{x})$ for a given model.

\begin{figure}
    \centering
    \includegraphics[width=0.5\textwidth]{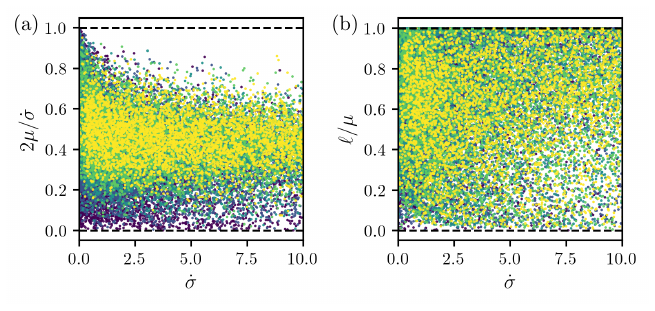}
    \caption{(a) Numerical results (color dots) for the optimized bound $\mu$. Dashed lines show \cref{eq:response-tur-opt}. (b) Numerical results (color dots) for the ratio $\ell/\mu$.  
    For the details of the simulations and description of colours see the caption of \cref{fig:fig-1}.
    }
    \label{fig:fig-2}
\end{figure}

\textit{Bound optimization.---}We notice that \cref{eq:response-tur-Markov} holds for an arbitrary linear combination of $j_e$ and an arbitrary dependence on $\varepsilon$, which implies that both $\boldsymbol{x}$ and $\boldsymbol{b}$ are arbitrary vectors. Therefore, for a given graph $\mathcal{G}$ and rate matrix $\mathbb{W}$, the optimized version of \cref{eq:response-tur-Markov} reads
\begin{align}
\label{eq:response-tur-opt}
    \mu \equiv \max_{\boldsymbol{\beta}} \max_{\boldsymbol{x}} m(\boldsymbol{\beta},\boldsymbol{x})&\leq \frac{\dot{\sigma}}{2}.
\end{align}
The problem $\max_{\boldsymbol{x}}m(\boldsymbol{\beta},\boldsymbol{x})$ is a standard convex optimization of a quadratic form, and has been previously applied to TUR~\eqref{eq:tur} in Ref.~\cite{vu2020entropy}. The optimal vector $\boldsymbol{x}^*$ reads
\begin{align}
\label{eq:opt-x}
    \boldsymbol{x}^*(\boldsymbol{\beta}) = (d_\varepsilon \boldsymbol{j}^\intercal \mathbb{C}^{-1} d_\varepsilon \boldsymbol{j})^{-1}\mathbb{C}^{-1}d_\varepsilon \boldsymbol{j}\,,
\end{align}
which results in $m(\boldsymbol{\beta},\boldsymbol{x}^*)$ for the inner maximum in \cref{eq:response-tur-opt}. To account for possible zero eigenvalues, the matrix $\mathbb{C}^{-1}$ is here the Drazin inverse~\cite{stanimirovic2014minimization}, which for real symmetric (and generally Hermitian) matrices is equivalent to Moore–Penrose pseudoinverse~\cite{landi2023current}. However, as shown in \cref{fig:fig-1}b, the optimization \cref{eq:opt-x} does not greatly improve the tightness of the bound \cref{eq:response-tur-Markov}. To do this, we also need to optimize $\boldsymbol{\beta}$.

The problem $\max_{\boldsymbol{\beta}}m(\boldsymbol{\beta},\boldsymbol{x})$ for a fixed $\boldsymbol{x}$ is solved simply by $\beta_e^*=\text{sign} (d_{B_e} \mathcal{J})$ due to the constraint $|\beta_e| \leq 1$. Therefore, the optimal vector $\boldsymbol{\beta}^*$ is a certain combination of $\pm 1$ elements. Thus, joint optimization over $\boldsymbol{\beta}$ and $\boldsymbol{x}$ can be performed by maximizing $m(\boldsymbol{\beta},\boldsymbol{x}^*)$ over all possible vectors $\boldsymbol{\beta}$ being combinations of $\pm 1$ elements. As shown in \cref{fig:fig-2}a in this case the bound \eqref{eq:response-tur-opt} can become tight, especially close to equilibrium, i.e., for small values of the entropy production.

\textit{Relation between R-TUR and TUR.---}We now prove that the conjectured R-TUR~\eqref{eq:response-tur} cannot be derived from the standard TUR~\eqref{eq:tur}.  Instead, it implies the validity of TUR. To that end, we recall the formulation of R-TUR via \cref{eq:response-tur-Markov} (which is equivalent to \cref{eq:response-tur}) and show that the standard TUR is equivalent to a weaker version of that bound,
\begin{align}
    \label{eq:response-tur-Markov-weak}
  \forall_{\boldsymbol{x}}: \quad m(\boldsymbol{1},\boldsymbol{x})
    = \frac{\big(\boldsymbol{1}\nabla\mathcal{J}\big)^2}{\langle \langle \mathcal{J} \rangle \rangle}&\leq \frac{\dot{\sigma}}{2}\,,
\end{align}
where $\boldsymbol{\beta}=\boldsymbol{1}$ (vector of ones) corresponds to homogeneous perturbation of the kinetic rates. This can be proven by noting that
\begin{align}
\boldsymbol{1} \nabla\mathcal{J} = \sum_{e} d_{B_e} \mathcal{J}=\sum_{e'} x_{e'} \sum_{e} d_{B_e} j_{e'}=\sum_{e'} x_{e'} j_{e'}=\mathcal{J}.
\end{align}
Here, in the second-last step, we used
\begin{align}
\label{eq:SRR}
    \sum_{e} d_{B_{e}} j_{e'} = j_{e'}\,.
\end{align}

which is the special case of the summation response relations introduced in Ref.~\cite{aslyamov2024general}: using $\partial_{B_e}j_e = j_e$ with 
\footnote{Eqs.~(11) and (13) of \cite{aslyamov2024general} }
we arrive at \cref{eq:SRR}. Inserting $\boldsymbol{1} \nabla\mathcal{J}=\mathcal{J}$ into \cref{eq:response-tur-Markov-weak} we obtain the standard TUR~\eqref{eq:tur}. Since \cref{eq:response-tur-Markov-weak} is a weaker version of \cref{eq:response-tur-Markov}, R-TUR implies TUR, while TUR does not necessarily imply R-TUR. Consequently, in the generic case, R-TUR cannot be derived from TUR.

\textit{Quantitative comparison with TUR.---}We now show that for most Markov networks \cref{eq:response-tur-Markov-weak} is indeed a weaker condition than \cref{eq:response-tur-Markov}. Consequently, for a given Markov network, the optimal response precision $(d_\varepsilon \mathcal{J})^2/\langle \langle \mathcal{J} \rangle \rangle$ can be greater than the optimal current precision $\mathcal{J}^2/\langle \langle \mathcal{J} \rangle \rangle$. As shown in Ref.~\cite{vu2020entropy}, the latter can be maximized as
\begin{align} \label{eq:tur-opt}
    \ell \equiv \max_{\boldsymbol{x}} \frac{\mathcal{J}^2}{\langle\langle\mathcal{J}\rangle\rangle}= m(\boldsymbol{1},\boldsymbol{x}^*) \geq \frac{\mathcal{J}^2}{\langle\langle\mathcal{J}\rangle\rangle}\,,
\end{align}
where \cref{eq:tur} also implies $\ell\leq \dot{\sigma}/2$; here we use the above proven relation $m(\boldsymbol{1},\boldsymbol{x})=\mathcal{J}^2/\langle \langle \mathcal{J} \rangle \rangle$. We thus obtain the chain of inequalities
\begin{align} \label{eq:chainineq}
    \frac{\mathcal{J}^2}{\langle\langle\mathcal{J}\rangle\rangle} \leq \ell \leq \mu \leq \frac{\dot{\sigma}}{2}\,,
\end{align}
with the equality $\ell =\mu$ for $\boldsymbol{\beta}^* =  \boldsymbol{1}$ (i.e., when the optimal $\boldsymbol{\beta}$ corresponds to homogeneous perturbation). This implies that the optimized R-TUR~\eqref{eq:response-tur-opt} provides a tighter bound for the entropy production than the standard TUR. We demonstrate this in \cref{fig:fig-2}b. Indeed, for certain models the ratio $\ell/\mu$ is close to 0, which implies that the optimized R-TUR~\eqref{eq:response-tur-opt} is far tighter than the standard TUR. Only in less than 10\% of the instances $\ell =\mu$, i.e., R-TUR~\eqref{eq:response-tur-opt} is optimized for homogeneous perturbation, which makes it equivalent to the optimized TUR~\eqref{eq:tur-opt}.

\begin{figure}
    \centering
    \includegraphics[width=0.98\linewidth]{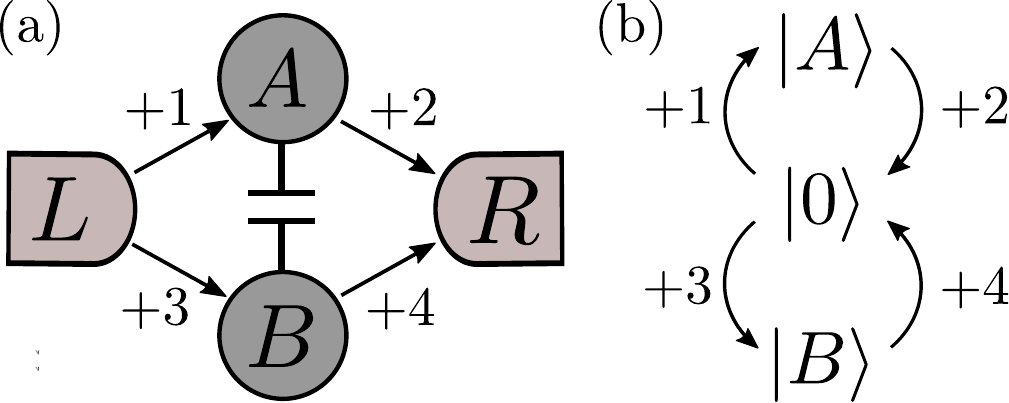}
    \caption{Scheme of the dynamical channel blockade model (a) and the corresponding Markov network (b).}
    \label{fig:fig-3}
\end{figure}

\textit{Example: dynamical channel blockade.---} To provide intuition why R-TUR can be tighter than the standard TUR, let us consider a minimal model where this occurs. 
It corresponds to the dynamical channel blockade model, which has been first considered in the context of electronic transport~\cite{bulka2000current, belzig2005full}, but is also similar to certain enzymatic reaction schemes~\cite{falasco2019negative}. 
The setup consists of two Coulomb-coupled quantum dots $A$ and $B$ attached to two electrodes $L$ and $R$ (Fig.~\ref{fig:fig-3}a). 
The system is subjected to a symmetric voltage $V>0$, with the chemical potentials of the electrodes parameterized as $\mu_L=-\mu_R=V/2$ and the dot energies set to $0$. Its dynamics is described by a Markov network (Fig.~\ref{fig:fig-3}b) consisting of three states $\{|0 \rangle,|A \rangle,|B \rangle \}$, corresponding the absence of electrons in the dots ($|0 \rangle$) and a single electron occupying the dots $A$ or $B$ (the double occupancy of the system is prohibited by the Coulomb coupling between dots). 
The edges of the Markov network $e \in \{1,2,3,4 \}$ correspond to electron jumps between the electrodes and the dots (Fig.~\ref{fig:fig-3}a). 
The transition rates read as $W_{\pm e}=\Gamma_e/\{ 1+\exp[ \mp V/(2 k_B T)]\}$, where $\Gamma_e$ are the tunneling rates, and their parameterization~\eqref{eq:rates-model} can be obtained by using $B_e = \ln \sqrt{W_{+e} W_{-e}}$ and $S_e=\ln(W_{+e}/W_{-e})=V/(2 k_B T)$.

\begin{figure}
    \centering
\includegraphics[width=0.5\textwidth]{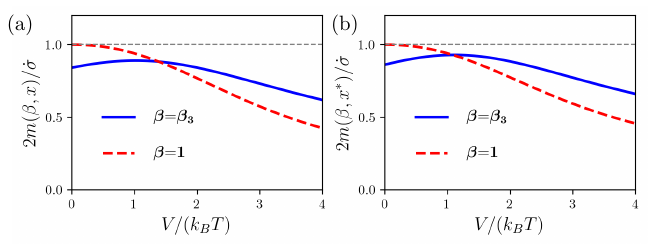}
    \caption{The coefficient $m(\boldsymbol{\beta},\boldsymbol{x})$ as a function of voltage, calculated for the particle current through the system (a) and the current optimized by \cref{eq:opt-x} (b). 
    The used vectors $\boldsymbol{\beta}=\boldsymbol{1} \equiv (1,1,1,1)^T$ and $\boldsymbol{\beta}=\boldsymbol{\beta}_3 \equiv (1,1,-1,1)^T$ are denoted by the labels in the plot. Parameters: $\Gamma_1=\Gamma_2=\Gamma_3=1$ , $\Gamma_4=0.2.$}
    
    \label{fig:fig-4}
\end{figure}

Let us now take the tunneling rate $\Gamma_4$ to be much smaller than the other ones. We first focus on the behavior of the coefficient $m(\boldsymbol{\beta},\boldsymbol{x})$ calculated for the most experimentally relevant current observable, namely, the particle current through the system $\mathcal{J}=j_1+j_3$ (which corresponds to $\boldsymbol{x}=(1,0,1,0)^T)$.  As shown in Fig.~\ref{fig:fig-4}a, for a large enough voltage (i.e., far of equilibrium), this coefficient is optimized for a vector $\boldsymbol{\beta}_3 \equiv (1,1,-1,1)^T$ rather than the uniform perturbation $\boldsymbol{\beta}=\boldsymbol{1}$. As discussed before, in such a case R-TUR provides a tighter bound for the entropy production than the standard TUR. To explain that, we recall that 
$m(\boldsymbol{\beta},\boldsymbol{x})$ is maximized for a vector $\boldsymbol{\beta}^*=(\ldots, \text{sign} (d_{B_e} \mathcal{J}) ,\ldots )^T$. For small $\Gamma_4$, after the transition $+3$, the electron gets stuck in the dot $B$ for a long time, blocking the electron transport. As a result, by suppressing the transition $3$, one suppresses the generation of the blocking state $|B \rangle$, and thus enhances the current; consequently, $d_{B_3} \mathcal{J}<0$. In contrast, the enhancement of other rates enhances the current ($d_{B_{1,2,4}} \mathcal{J}>0$). Thus, $\boldsymbol{\beta}^*=\boldsymbol{\beta}_3$. The same conclusion holds for the optimized current observable corresponding to the vector $\boldsymbol{x}^*$ given by \cref{eq:opt-x} (Fig.~\ref{fig:fig-4}b), which is here very close to the particle current considered before.

\textit{Proof for local responses.---}We have proven that in general R-TUR cannot be derived from TUR. However, this is possible for some specific cases. The first example is the response of a single edge current $\mathcal{J} = j_e$ to the perturbation of that edge. Applying the results of Ref.~\cite{aslyamov2024general} to the rate parameterization \eqref{eq:rates-model} (which implies $\partial_{B_e}j_e = j_e$), we arrive at
\begin{align} \label{eq:bound-loc}
0 \leq j_e^{-1} d_{B_e} j_e \leq 1.
\end{align}
In conjunction with TUR~\eqref{eq:tur}, this implies Eq.~\eqref{eq:response-tur_Simplif}:
\begin{align} \label{eq:local-proof}
\frac{(d_{B_e} j_e)^2}{\langle\langle j_e \rangle\rangle}  \leq  \frac{ j_e^2}{\langle\langle j_e \rangle\rangle}
 \leq \frac{\dot{\sigma}}{2}\,.
\end{align}

\textit{Proof for unicyclic networks.---}\cref{eq:response-tur} can also be proven for unicyclic networks, i.e., systems in which each state $i \in \{0,\ldots,N-1\}$ is a source of a single directed edge pointing to the tip $i+1$ (with $i$ defined modulo $N$). In such a case, all edge currents $j_e=j$ are equal to each other and $\mathcal{J} =\sum_e x_e j_e=x_\Sigma j$, where $x_\Sigma=\sum_e x_e$. 
Moreover, we have
\begin{align}
\label{eq:sign-one-cycle}
    \text{sign}\,d_{B_e}\mathcal{J} = \text{sign}\,x_\Sigma d_{B_e}j = \text{sign}\,x_\Sigma j\,,
\end{align}
where we use \cref{eq:bound-loc} for $d_{B_e}j/j\geq 0$. We can always define the orientation of $\mathcal{G}$ so that $x_\Sigma j$ is positive, which implies $\beta_e^*=\text{sign}(d_{B_e}\mathcal{J})=1$ (see \cref{eq:sign-one-cycle}), i.e., $\mu$ is maximized for homogeneous perturbation. Then, as discussed below \cref{eq:chainineq}, $\mu=\ell \leq \dot{\sigma}/2$, which proves \cref{eq:response-tur}.

\textit{Proof for the linear response regime.---}Finally, R-TUR~\eqref{eq:response-tur} can be proved to hold close to equilibrium, where the currents are linear in the applied thermodynamic forces. To define this regime, let us parameterize the asymmetric transition rate parameters in \cref{eq:rates-model} as $S_e=F_e+E_{s(e)}-E_{t(e)}$~\cite{owen2020universal,rao2018conservation}, where $s(e)$ and $t(e)$ are the source and the tip of the edge $e$, $E_i$ are node parameters (which may sometimes be interpreted as state energies), and $F_e$ are the nonconservative thermodynamic forces. In the absence of forces, the system relaxes to the equilibrium state $\pi_i^\text{eq} =e^{-E_i}/Z$, with $Z=\sum_i e^{-E_i}$, which is independent of $B_e$. Close to equilibrium (i.e., for small values of nonconservative forces) the currents respond linearly to applied forces as $\boldsymbol{j}=\mathbb{L} \boldsymbol{F}$, where $\boldsymbol{F}=(\ldots,F_e,\ldots)^T$ and $\mathbb{L}$ is the positive semidefinite Onsager matrix. Consequently, the current observables and the entropy production can be calculated as $\mathcal{J} = \boldsymbol{x}^T \mathbb{L} \boldsymbol{F}$ and $\dot{\sigma}=\boldsymbol{F}^T \mathbb{L} \boldsymbol{F}$. Furthermore, in this regime, the covariance matrix of edge currents is related to Onsager matrix as $\mathbb{C}=2 \mathbb{L}$ by virtue of the fluctuation-dissipation theorem~\cite{vroylandt2019ordered}. Let us also define the matrix $\mathbb{L}_{\boldsymbol{\beta}} \equiv d_\varepsilon \mathbb{L}/b_\text{max}=\sum_e \beta_e d_{B_e} \mathbb{L}$, which describes the kinetic response of Onsager matrix. Inserting the above expressions for $\mathbb{C}$ and $\dot{\sigma}$ into \cref{eq:response-tur-Markov}, using $\boldsymbol{\beta}\nabla\mathcal{J}=\sum_e \beta_e d_{B_e} \mathcal{J}=\boldsymbol{x}^T \mathbb{L}_{\boldsymbol{\beta}} \boldsymbol{F}$, and applying optimization~\eqref{eq:opt-x}, we find that R-TUR~\eqref{eq:response-tur} holds if and only if
\begin{align}
\forall_{\boldsymbol{F}}: \quad \boldsymbol{F}^T \mathbb{L}_{\boldsymbol{\beta}} \mathbb{L}^{-1} \mathbb{L}_{\boldsymbol{\beta}} \boldsymbol{F} \leq  \boldsymbol{F}^T \mathbb{L} \boldsymbol{F} \,.
\end{align}
We recall that $\mathbb{L}^{-1}$ here denotes the Moore-Penrose pseudoinverse of $\mathbb{L}$. This is equivalent to
\begin{align} \label{eq:rtur-lin-semipos}
\mathbb{L}-\mathbb{L}_{\boldsymbol{\beta}} \mathbb{L}^{-1} \mathbb{L}_{\boldsymbol{\beta}} \succcurlyeq 0 \,,
\end{align}
where $\mathbb{A} \succcurlyeq 0$ means that the matrix $\mathbb{A}$ is positive semidefinite. To prove the above inequality, we note that $\mathbb{L}-\mathbb{L}_{\boldsymbol{\beta}} \mathbb{L}^{-1} \mathbb{L}_{\boldsymbol{\beta}}$ is the Schur complement $\mathbb{M}/\mathbb{L}$ of the matrix
\begin{align}
\mathbb{M} = \begin{pmatrix} \mathbb{L} & \mathbb{L}_{\boldsymbol{\beta}} \\ \mathbb{L}_{\boldsymbol{\beta}} & \mathbb{L} 
\end{pmatrix} \,.
\end{align}
Thus, \cref{eq:rtur-lin-semipos} holds if $\mathbb{M}  \succcurlyeq 0$. Since eigenvalues of $\mathbb{M}$ correspond to that of $\mathbb{L} \pm \mathbb{L}_{\boldsymbol{\beta}}$~\cite{nath2014distance}, $\mathbb{M} \succcurlyeq 0$ if and only if $\mathbb{L} \pm \mathbb{L}_{\boldsymbol{\beta}} \succcurlyeq 0$. To prove the latter occurrence, in Appendix~\ref{sec:semipos-onsager-proof} we show that the responses $d_{B_e} \mathbb{L}$ are positive semidefinite and obey the summation response relation
\begin{align} \label{eq:onsager-summation}
\sum_e d_{B_e}\mathbb{L}=\mathbb{L} \,.
\end{align}
As a consequence, $\mathbb{L} \pm \mathbb{L}_{\boldsymbol{\beta}}=\sum_e (1 \pm\beta_e) d_{B_e} \mathbb{L}$ is positive semidefinite because $1 \pm \beta_e \in [0,2] \geq 0$ and $d_{B_e} \mathbb{L} \succcurlyeq 0$. This proves \cref{eq:rtur-lin-semipos}, and thus R-TUR~\eqref{eq:response-tur}.

We also note that \cref{eq:rtur-lin-semipos} is saturated for homogeneous perturbation $\boldsymbol{\beta}=\boldsymbol{1}$. Thus, in the linear response regime, the optimized R-TUR~\eqref{eq:response-tur-opt} is equivalent to the optimized TUR~\eqref{eq:tur-opt}. (We note that, in contrast to the unicyclic case, this does not imply that homogeneous perturbation will always yield the optimal response precision when considering a generic rather than optimized current observable.) Using \cref{eq:opt-x}, both are optimized for the current observable corresponding to the entropy production: $\mathcal{J}=\dot{\sigma}$. Indeed, in the linear response regime, $\langle \langle \dot{\sigma} \rangle \rangle=2 \dot{\sigma}$, and thus TUR~\eqref{eq:tur} becomes saturated: $\dot{\sigma}^2/\langle \langle \dot{\sigma} \rangle \rangle=\dot{\sigma}/2$~\cite{pietzonka2016universal}. Furthermore, using $d_{B_e} \mathbb{L} \succcurlyeq 0$, \cref{eq:onsager-summation} and $\dot{\sigma}=\boldsymbol{F}^T \mathbb{L} \boldsymbol{F}$, we find that close to equilibrium the entropy production response itself, and not only its precision, is bounded by dissipation as
\begin{align}
&0 \leq d_{B_e} \dot{\sigma} \leq \dot{\sigma} \quad \text{with} \quad \sum_e d_{B_e} \dot{\sigma} = \dot{\sigma} \,, \\
& |d_\varepsilon \dot{\sigma}|  \leq b_\text{max} \dot{\sigma} \,.
\end{align}

Finally, we note that the Onsager matrix can be represented in the basis of measurable currents (e.g., charge, heat, or chemical currents) conjugated to fundamental thermodynamic forces (e.g., temperature or chemical potential gradients)~\cite{forastiere2022linear}. Such physical Onsager matrix can be expressed as $\mathbb{L}_{\mathbb{X}}=\mathbb{X} \mathbb{L} \mathbb{X}^T$, where $\mathbb{X}$ is the matrix that connects measurable currents to edge currents~\cite{vroylandt2019ordered}. Using $d_{B_e} \mathbb{L} \succcurlyeq 0$ and \cref{eq:onsager-summation}, its kinetic response can also be bounded as
\begin{align}
&\mathbb{L}_{\mathbb{X}} \succcurlyeq d_{B_e} \mathbb{L}_{\mathbb{X}} \succcurlyeq 0 \quad \text{with} \quad \sum_e d_{B_e} \mathbb{L}_{\mathbb{X}}=\mathbb{L}_{\mathbb{X}} \,, \\
& b_\text{max} \mathbb{L}_{\mathbb{X}} \succcurlyeq d_\varepsilon \mathbb{L}_{\mathbb{X}} \succcurlyeq -b_\text{max} \mathbb{L}_{\mathbb{X}} \,.
\end{align}
This enables one to experimentally bound $b_\text{max}$ using only measurable currents. 

\textit{Final remarks.---}Our bound establishes a fundamental thermodynamic cost for controlling currents by modulating the system's kinetics. 
It may prove useful in optimizing the energetic cost needed to accurately control nanodevices such as molecular machines or nanoelectronic devices.
It may also be used for thermodynamic~\cite{seifert2019stochastic} or model~\cite{moffitt2010methods} inference. In particular, when all parameters $b_e$ are known (e.g., when the control parameter $\varepsilon$ affects only some known transition rates), the empirical current responses and fluctuations may provide a better estimate of the entropy production than the standard TUR. On the other hand, when the model details are unknown but the entropy production is accessible,~\cref{eq:response-tur} can be used to bound the transition rate responses $b_e$.

\begin{acknowledgments}

K.P., T.A. and M.E. acknowledge the financial support from, repectively, project No.\ 2023/51/D/ST3/01203 funded by the National Science Centre, Poland, project ThermoElectroChem (C23/MS/18060819) from Fonds National de la Recherche-FNR, Luxembourg, project TheCirco (INTER/FNRS/20/15074473) funded by FRS-FNRS (Belgium) and FNR (Luxembourg).
\end{acknowledgments}

\appendix
\section{Numerical Simulations} \label{sec:numerical}

To calculate the response $\nabla\mathcal{J}$ we use the results of Ref.~\cite{aslyamov2024general} as follows
\begin{align}
\label{eq:appendix-response}
    \nabla^\intercal\mathcal{J} =\Big[\sum_e x_e d_{B_{e'}}j_e \Big]_{\{e'\in\mathcal{E}\}}=\boldsymbol{x}^\intercal\mathbb{P}\mathbb{J}\,,
\end{align}
where we use notation $[f_e]_{\{e\in\mathcal{E}\}}=(\dots,f_e,\dots)^\intercal$ for the vector $\boldsymbol{f}$ and where 
the matrix $\mathbb{P}$ is the projection matrix with analytical expression \cite{aslyamov2024general}, $\mathbb{J}=[\partial_{B_{e'}}j_e]_{\{e\in\mathcal{E},e'\in\mathcal{E}\}}=\text{diag}(\dots,j_e,\dots)$ the diagonal Jacobian matrix. 

We find $\mathbb{P}$ from Eq.~(8) of \cite{aslyamov2024general} (which demands the knowledge of rates $W_e$ and $\boldsymbol{\pi}$) and find $\mathbb{J}$ from $j_e = W_{+e}\pi_{s(+e)}-W_{-e}\pi_{s(-e)}$.
Thus, to simulate \cref{eq:appendix-response}, we only need the model for the rates $W_{\pm e}$ and the numerical values of the steady state $\boldsymbol{\pi}$ satisfying \cref{eq:NESS}. 

To find the covariance matrix of the edge currents $\mathbb{C}$, we follow the method from Ref.~\cite{wachtel2015fluctuating} and define the matrix $\mathbb{W}^\phi(\boldsymbol{q})$ with nondiagonal elements $W^\phi_{t(\pm e)s(\pm e)}(\boldsymbol{q})=W_{t(\pm e)s(\pm e)}\exp(\pm q_e)$ and diagonal elements the same as $\mathbb{W}$. The elements of the covariance matrix are defined as
\begin{align}
\label{eq:appendix-cov-mat}
    C_{ee'} = \frac{\partial}{\partial q_e}\frac{\partial}{\partial q_{e'}} \lambda(\boldsymbol{q})\Big|_{\boldsymbol{q}=\boldsymbol{0}}\,,
\end{align}
where $\lambda$ is eigenvalue of the matrix $\mathbb{W}^\phi$ with the largest real part. In simulations of \cref{eq:appendix-cov-mat}, we used the method of finite differences to numerically calculate $\lambda(\boldsymbol{q})$.

\section{Proof of $d_{B_e} \mathbb{L} \succcurlyeq 0$ and Eq.~\eqref{eq:onsager-summation}} \label{sec:semipos-onsager-proof}
To prove $d_{B_e} \mathbb{L} \succcurlyeq 0$ we use the expression for Onsager matrix from Ref.~\cite{vroylandt2019ordered},
\begin{align} \label{eq:onsager}
\mathbb{L}=\tfrac{1}{2} \mathbb{K} \mathbb{H}^{-1} \mathbb{K}^T \quad \text{with} \quad \mathbb{H}=\mathbb{K}^T \mathbb{T}^{-1} \mathbb{K} \,.
\end{align}
Here $\mathbb{K}$ is the cycle matrix that depends only on the system topology and not on transition rates, while $\mathbb{T}=\text{diag}(\ldots,\tau_e^\text{eq},\ldots)$ is the matrix with equilibrium traffics $\mathcal{\tau}_e^\text{eq}=W_{+e} \pi^\text{eq}_{s(+e)}+W_{-e} \pi^\text{eq}_{s(-e)}$ at the diagonal. We note that Ref.~\cite{vroylandt2019ordered} actually considered the Onsager matrix describing current responses to edge affinities $\mathcal{F}_e$ rather than forces $F_e$ defined here; however, the Onsager matrix for both force definitions is the same, as it is related to the covariance matrix via the fluctuation-dissipation theorem~\cite{pietzonka2016universal} $\langle \langle \dot{\sigma} \rangle \rangle =2 \dot{\sigma} \Leftrightarrow \mathbb{L}=\mathbb{C}/2$. We now use the expressions $d_{B_e} \mathbb{H}^{-1}=-\mathbb{H}^{-1} (d_{B_e} \mathbb{H}) \mathbb{H}^{-1}$, and the equality $d_{B_e} \tau_{e'}^\text{eq}=\delta_{ee'} \tau_e^\text{eq}$ that holds because probabilities $\pi_i^\text{eq}$ do not depend on $B_e$; the latter yields $d_{B_e} \mathbb{T}^{-1}=-\mathbb{T}_e^{-1}$ with $\mathbb{T}_e=\text{diag}(\ldots,0,\tau_e^\text{eq},0,\ldots)$. We get
\begin{align} \label{eq:onsager-response}
d_{B_e} \mathbb{L}=\tfrac{1}{2} \mathbb{K} \mathbb{H}^{-1} \mathbb{K}^T \mathbb{T}_e^{-1} \mathbb{K} \mathbb{H}^{-1} \mathbb{K}^T=2 \mathbb{L} \mathbb{T}_e^{-1} \mathbb{L} \,.
\end{align}
This is a positive semidefinite matrix, as this is a sandwich of the positive semidefinite diagonal matrix $\mathbb{T}_e^{-1}$ between the matrix $\mathbb{L}$ and its transpose $\mathbb{L}^T=\mathbb{L}$, which concludes the proof of $d_{B_e} \mathbb{L} \succcurlyeq 0$. 

Summing \cref{eq:onsager-response} over edges, we further obtain \cref{eq:onsager-summation}:
\begin{align} \label{eq:onsager-summation-proof}
\sum_e d_{B_e}\mathbb{L}=\tfrac{1}{2} \mathbb{K} \mathbb{H}^{-1} \mathbb{H} \mathbb{H}^{-1} \mathbb{K}^T=\mathbb{L} \,.
\end{align}
Here we use $\sum_e \mathbb{T}_e^{-1} = \mathbb{T}^{-1}$ and the relation $\mathbb{H}^{-1} \mathbb{H} \mathbb{H}^{-1}=\mathbb{H}^{-1}$ that holds for Moore-Penrose pseudoinverse.


\bibliography{biblio}
\end{document}